\documentclass[runningheads]{llncs}
\usepackage[T1]{fontenc}
\usepackage{graphicx}
\usepackage{algpseudocode}
\usepackage{algorithm}
\usepackage{booktabs, multirow}
\usepackage{csquotes}
\usepackage{tikz}
\usetikzlibrary{positioning,fit}
\usepackage{float}

\begin{document}
\title{Modeling Social Media Recommendation Impacts Using Academic Networks: A Graph Neural Network Approach}
\titlerunning{Modeling Social Media Recommendation Impacts Using Academic Networks}
%
\author{Sabrina Guidotti\inst{1} \and Gregor Donabauer\inst{2} \and Simone Somazzi\inst{1} \and Udo Kruschwitz\inst{2} \and Davide Taibi\inst{3} \and Dimitri Ognibene\inst{1}}
\authorrunning{S. Guidotti et al.}
%
\institute{
Università degli Studi di Milano-Bicocca, Milan, Italy \and
University of Regensburg, Regensburg, Germany \and
National Research Council of Italy, Palermo, Italy \\
\email{dimitri.ognibene@unimib.it}
}

\maketitle              
\begin{abstract}

The widespread use of social media has highlighted potential negative impacts on society and individuals, largely driven by recommendation algorithms that shape user behavior and social dynamics. Understanding these algorithms is essential but challenging due to the complex, distributed nature of social media networks as well as limited access to real-world data. This study proposes to use academic social networks as a proxy for investigating recommendation systems in social media. By employing Graph Neural Networks (GNNs), we develop a model that separates the prediction of academic infosphere from behavior prediction, allowing us to simulate recommender-generated infospheres and assess the model's performance in predicting future co-authorships. Our approach aims to improve our understanding of recommendation systems' roles and social networks modeling. To support the reproducibility of our work we publicly make available our implementations: \url{https://github.com/DimNeuroLab/academic\_network\_project}

\keywords{Social Networks \and Societal Well-Being \and GNNs.}
\end{abstract}

\section{Introduction}

The widespread use of social media has revealed various aspects that may negatively affect both society and individuals using such platforms \cite{almourad2020defining,gillani2018me,marengo2018highly}. Social networks are complex systems where user interactions are heavily influenced by recommendation algorithms, which shape user behavior and, in turn, the broader social impact \cite{ognibene2023challenging}. Examples for algorithmic threats on social media include \textit{filter bubbles} \cite{nikolov2015measuring} and \textit{echo chambers} \cite{gillani2018me}. Understanding the underlying mechanisms is crucial for mitigating negative effects and promoting societal well-being \cite{ognibene2023challenging}. However, studying these mechanisms is challenging \cite{chen2018survey,eirinaki2018recommender} due to the distributed, heterogeneous, and large-scale nature of social media networks \cite{eksombatchai2018pixie,covington2016deep} as well as a lack of access to real-world data, making it difficult to align theoretical models with real-world dynamics.

Academic social networks, which share similarities with social media networks \cite{kong2019academic,jordan2019social}, could serve as a proxy for research in the field \cite{fortunato2018science}. They provide easier access to data \cite{tang2008arnetminer} and allow to study connectivity dynamics when recommendation systems are introduced. However, modeling academic behavior is complex due to the non-local nature of academic activities, which involve diverse topics, venues, and collaborations \cite{kong2019academic}. This requires analyzing a large set of potential interactions, often overlooking computational efficiencies that could be achieved through network structures. Such activities are likely determined by an academic being exposed to non-local information presented to them by recommenders (infosphere) \cite{chen2023bias}. Characterizing the recommenders' contribution could therefore lead to more realistic academic models and improve computational efficiency.

In this work, we aim to better understand the role of recommendation systems in social media by using academic networks as a proxy. We employ Graph Neural Networks (GNNs) to develop a model that separates predicting an academic infosphere from predicting its behavior. This allows us to simulate a potential infosphere generated by a recommender system and evaluate the model’s performance in link prediction of future co-authorships.

\section{Related Work}

Much research in recommender systems focuses on the broader social and behavioral impact. Studies for example showed that the presentation of recommendations can significantly influence user satisfaction \cite{nanou2010effects} or that social explanations can affect user interactions but do not always improve satisfaction with the content \cite{sharma2013social}. Recommenders are also known to influence consumer preferences through mechanisms like the anchoring effect when users trust the system \cite{adomavicius2013recommender}.

Simulations of recommenders can help to reveal effects of such systems. They have for example investigated how repeated interactions can amplify biases, such as popularity bias and filter bubbles, impacting long-term user behavior \cite{yao2021measuring} or that personalized recommendations can increase commonality among users leading to increased consumption and a more homogeneous product mix \cite{fleder2010recommender,hosanagar2014will}.

For example, addressing biases in recommenders is crucial, as they can lead to discrepancies between offline evaluations and online performance, negatively affecting user trust which requires efforts in debiasing such systems \cite{chen2023bias} and understanding how biases propagate through user profiles can help in developing more accurate and fair recommendation algorithms \cite{lin2020calibration}.

Building on work that has used academic networks in tasks like web user profiling \cite{Tang:10TKDD}, topic expertise search \cite{Tang:11ML} or social network extraction of academics \cite{Tang:07ICDM}, we want to use such data to improve understanding how different simulated infospheres influence learning of user behaviour.

\section{Methodology}

Our approach involves three key components: \textbf{(1)} characterizing the \textbf{history} of the agents' behavior within the social network, \textbf{(2)} simulating a recommender system by integrating \textbf{infosphere} into the network, and \textbf{(3)} learning models on these networks to evaluate how different infospheres influence the model's prediction performance of the agents' future behavior.

\subsection{Agent History and Infosphere Simulation}

For modelling the authors' \textbf{History} we track the papers they have written, the co-authors they have collaborated with, and the topics they have worked on. This history is represented as a time-dependent graph $G_{y}$, containing all observable information up to year $y$. Each graph $G_{y}$ includes three types of nodes: \textit{author}, \textit{paper}, and \textit{topic}, as well as three types of edges: (\textit{author, writes, paper}), (\textit{paper, deals\_with, topic}), and (\textit{paper, cites, paper}). Paper nodes are characterized by the year of publication. All other nodes in the graph are initialized with random embeddings which are jointly learnt during model training.

We then introduce the concept of \textbf{Infosphere}, representing the information an author might encounter through a recommender system or other algorithmic components. Such systems access content and authors beyond the author's direct and local experience, with suggestions guided by the author's prior interactions rather than random selection. Our concept of infosphere is related to the idea of \textit{impressions}, which are defined as the recommendations presented to the user along with their corresponding interactions \cite{perez_et_al_2020}. Although difficult to model due to limited knowledge of the underlying algorithms, we simulate the infosphere with minimal assumptions, ensuring that it includes both information the author has definitely encountered and what they might have encountered through similar processes. We use information from subsequent years' graphs to derive a set of paths that represent the author's connections with nodes in their history as they appear in the following year within the current year's graph. By adding noise based on a set of probabilities which determine whether to follow existing paths or switch directions we aim to make the simulation more realistic. This approach allows us to develop models that keep user and recommender modeling separate, so that the recommenders can pursue different objectives. By creating a minimal infosphere, we develop a more robust and general model of authors, which can be used to refine recommender systems based on principles such as content similarity. For comparison, we also generated alternative infospheres: one consisting of the $n$ most popular papers in a given year $y$, and another focusing on the $n$ most popular papers within the $m$ most-used topics by an author $x$ in that year.

\subsection{Seedgraph and Expansion}

Our infosphere calculation is based on a seedgraph, a directed graph composed of paths associated with each author. Each path traces the shortest connection from an element in the author's history in year $y+1$ back to the graph in year $y$.

We first initialize a "frontier-seeds" dictionary with the author's publications and related information, then iteratively expand both the author node and the "frontier-seeds" by 1-hop. A "compare-frontiers" function identifies paths when overlaps occur. The process continues until all paths are identified, allowing efficient seedgraph construction by merging node lists when common nodes are found.

To achieve realistic expansion, the seedgraph is extended with plausible alternative paths. For better understanding, nodes can be interpreted as colored as white, orange, or green, where orange nodes belong to the seed graph, green nodes are added during expansion, and white nodes belong to neither. The algorithm requires several inputs: the author-node, the full-graph, the seed graph, and parameters (p1, p2, p3, f):

\begin{itemize}
    \item[\textbf{p1:}] Probability of following a path of orange nodes (seedgraph). Higher values extend the original infosphere.
    \item[\textbf{p2:}] Probability of following a path of green nodes (expanded graph). Higher values create paths similar to the seedgraph.
    \item[\textbf{p3:}] Probability of returning to the author node. Higher values concentrate noise near the author node.
    \item[\textbf{f:}] Number of new nodes added per seedgraph path (2, 4, or 6).
\end{itemize}


\subsection{Behaviour Prediction}

To assess the impact of the simulated recommender via infosphere on the prediction of future actions of agents (authors) within the social network, we evaluate how the configurations described previously affect a model's ability to forecast future co-authorship collaborations.

We model the task of co-authorship prediction as a link prediction problem and learn a GNN that predicts future co-authorships that do not yet exist as edges in the current graph. To do this, we add co-author edges derived by multi-hop walks from author nodes across the existing heterogeneous graph. In both scenarios - predicting links within a specific year or forecasting future connections - we also add and sample negative examples of edges at a ratio of 1:1.

\section{Experiments}

\subsection{Dataset}

We use the \textit{DBLP-Citation-network v14} dataset \cite{tang2008arnetminer} from AMiner, which includes data from sources like DBLP, ACM, and MAG to provide a comprehensive overview of academic publications and their citation relationships. We selected this dataset as it is the most up-to-date dataset available (released in 2023) and offers a reasonable number of nodes and edges making it a good fit for our experiments. Specifically, it contains 5,259,858 paper nodes and 36,630,661 citation edges. Additionally, each paper is associated with related information, such as authors, venues, and topics, which can be modeled as additional nodes in a heterogeneous network.

\subsection{Infosphere Parameters}
\label{subsec:inf_parameters}

We evaluate different parameters for creating the infosphere as described previously. When reporting results based on these combinations we refer to the run ids assigned here.

\begin{itemize}
    \item \textit{trial0} Random Infosphere
    \item \textit{trial1} (p1=0.5; p2=0.5; p3=0.5; f=2)
    \item \textit{trial2} (p1=0.75; p2=0.5; p3=0.5; f=2)
    \item \textit{trial3} (p1=0.5; p2=0.75; p3=0.5; f=2)
    \item \textit{trial4} (p1=0.5; p2=0.5; p3=0.75; f=2)
    \item \textit{trial5} (p1=0.25; p2=0.75; p3=0.25; f=2)
\end{itemize}


    

\subsection{Model and Training}

For the learning process, we use an encoder-decoder network. The encoder consists of a heterogeneous Graph Neural Network with two consecutive graph convolution layers to encode the input graphs and generate expressive node representations. In our experiments we evaluated GraphSAGE \cite{hamilton2017inductive} and Heterogeneous Graph Transformer (HGT) \cite{hu2020heterogeneous} as encoder layers. While HGT uses Transformer blocks for neighborhood aggregation we vary the aggregation strategies for GraphSAGE. The decoder is a simple two-layer feed-forward neural network that classifies node pairs as either connected (existing edge/link) or not, making the task as a binary classification problem. The model is trained end-to-end using binary cross-entropy loss and the Adam optimizer. We run training for 500 epochs with early stopping and a patience of 10, the batch-size is set to $1024$ and the the learning rate to $0.00001$.

\section{Results}

Table \ref{table:results_co_author} summarizes the results from our experiments on predicting co-authors using various model setups and infosphere configurations. We achieved an accuracy of 78-80\% when predicting co-authors without any infosphere. When incorporating the infosphere generated by our methods, performance improved, with accuracy reaching around 88\%, particularly in setups without seedgraph expansions. Testing different aggregation functions (sum, min, mean, max) yielded minimal differences in outcomes.

We also evaluated an alternative infosphere based on the most popular papers (10 and 50). This approach performed worse than using the future infosphere and was comparable to or worse than having no infosphere, likely due to the noise introduced by less relevant papers. The min aggregation function slightly improved results but remained close to those without an infosphere.

\begin{table}[H]
\centering
\caption{Accuracy scores for prediction of next year's co-authors without and with different infospheres as well as various model setups. \textit{Infosphere dropped} refers to the number of infosphere edges that were removed in the respective setup. \textit{Infosphere Params} refers to the trials introduced in Section \ref{subsec:inf_parameters} for infosphere type author, note that some (2-4) are omitted due to similar results.}
\label{table:results_co_author}
\scriptsize
\begin{tabular}{|l|l|l|l|l|l|}
\hline
\textbf{Inf. Type} & \textbf{Inf. Params} & \textbf{Inf. Dropped} & \textbf{Accuracy} & \textbf{Aggregation} & \textbf{GNN Type} \\
\hline
- & - & - & 0.788 & max & SAGE \\
- & - & - & 0.780 & mean & SAGE \\
- & - & - & 0.796 & min & SAGE \\
- & - & - & 0.802 & sum & SAGE \\
- & - & - & 0.798 & N/A & HGT \\
\hline
author & 0 & - & 0.892 & max & SAGE \\
author & 0 & - & 0.892 & mean & SAGE \\
author & 0 & - & 0.893 & min & SAGE \\
author & 0 & - & 0.886 & sum & SAGE \\
author & 0 & - & 0.893 & N/A & HGT \\
\hline
author & 5 & - & 0.889 & max & SAGE \\
author & 5 & - & 0.891 & mean & SAGE \\
author & 5 & - & 0.890 & min & SAGE \\
author & 5 & - & 0.888 & sum & SAGE \\
\hline
author & 0 & 10\% & 0.888 & sum & SAGE \\
author & 0 & 25\% & 0.881 & sum & SAGE \\
author & 0 & 50\% & 0.883 & sum & SAGE \\
author & 0 & 75\% & 0.862 & sum & SAGE \\
author & 0 & 90\% & 0.834 & sum & SAGE \\
author & 0 & 100\% & 0.789 & sum & SAGE \\
\hline
top-paper & 10 & - & 0.501 & sum & SAGE \\
top-paper & 10 & - & 0.585 & max & SAGE \\
top-paper & 10 & - & 0.772 & mean & SAGE \\
top-paper & 10 & - & 0.750 & min & SAGE \\
top-paper & 10 & - & 0.508 & sum & SAGE \\
top-paper & 10 & - & 0.662 & N/A & HGT \\
\hline
top-paper & 50 & - & 0.649 & sum & SAGE \\
top-paper & 50 & - & 0.741 & max & SAGE \\
top-paper & 50 & - & 0.745 & mean & SAGE \\
top-paper & 50 & - & 0.793 & min & SAGE \\
top-paper & 50 & - & 0.704 & sum & SAGE \\
top-paper & 50 & - & 0.770 & N/A & HGT \\
\hline
top-paper-per-topic & [1,10] & - & 0.678 & sum & SAGE \\
top-paper-per-topic & [1,10] & - & 0.754 & max & SAGE \\
top-paper-per-topic & [1,10] & - & 0.690 & mean & SAGE \\
top-paper-per-topic & [1,10] & - & 0.748 & min & SAGE \\
top-paper-per-topic & [1,10] & - & 0.643 & sum & SAGE \\
top-paper-per-topic & [1,10] & - & 0.728 & N/A & HGT \\
\hline
top-paper-per-topic & [1,50] & - & 0.667 & sum & SAGE \\
top-paper-per-topic & [1,50] & - & 0.745 & max & SAGE \\
top-paper-per-topic & [1,50] & - & 0.676 & mean & SAGE \\
top-paper-per-topic & [1,50] & - & 0.789 & min & SAGE \\
top-paper-per-topic & [1,50] & - & 0.621 & sum & SAGE \\
top-paper-per-topic & [1,50] & - & 0.778 & N/A & HGT \\
\hline
top-paper-per-topic & [2,5] & - & 0.723 & sum & SAGE \\
top-paper-per-topic & [2,5] & - & 0.755 & N/A & HGT \\
\hline
top-paper-per-topic & [5,10] & - & 0.709 & sum & SAGE \\
top-paper-per-topic & [5,10] & - & 0.784 & N/A & HGT \\
\hline
top-paper-per-topic & [10,1] & - & 0.562 & sum & SAGE \\
top-paper-per-topic & [10,1] & - & 0.734 & N/A & HGT \\
\hline
top-paper-per-topic & [10,5] & - & 0.721 & sum & SAGE \\
top-paper-per-topic & [10,5] & - & 0.787 & N/A & HGT \\
\hline
top-paper-per-topic & [50,1] & - & 0.753 & sum & SAGE \\
top-paper-per-topic & [50,1] & - & 0.781 & N/A & HGT \\
\hline
\end{tabular}
\end{table}

Finally, we evaluated an infosphere configuration that connects authors to the most popular papers within their frequently used topics. Various combinations of topics and papers were evaluated and the results are again much lower than with author-based infosphere, sometimes even falling below the scores we get when using no infosphere at all.

\section{Conclusion}
\sloppy
In this study, we successfully demonstrated how simulating recommenders through infospheres helps understanding user behavior when based on academic network data. Our analysis showed that the infosphere most benefits predictions of new edges that are not present in the history. These findings can help to improve the understanding of how recommender mechanisms influence communities, especially those currently exposed to negative impacts in social networks.

\begin{credits}
\subsubsection{\ackname} We want to thank the anonymous reviewers for their insightful comments which have helped us to improve the paper.
\end{credits}

%
%
%
\bibliographystyle{splncs04}
\bibliography{main}
\end{document}